\documentclass[fleqn,usenatbib]{mnras}

\usepackage{newtxtext,newtxmath}

\usepackage[T1]{fontenc}

\DeclareRobustCommand{\VAN}[3]{#2}
\let\VANthebibliography\thebibliography
\def\thebibliography{\DeclareRobustCommand{\VAN}[3]{##3}\VANthebibliography}

\usepackage{graphicx}	
\usepackage{amsmath}	
\usepackage{subfigure, color}
\usepackage{xcolor}
\usepackage{cleveref}

\title[Galactic cosmic rays fluxes on M dwarfs]{Galactic cosmic ray propagation through M dwarf planetary systems}

\author[A. L. Mesquita et al.]{
A. L. Mesquita$^{1,2}$\thanks{E-mail: mesquita@tcd.ie},
D. Rodgers-Lee$^{1}$, 
A. A. Vidotto$^{1,2}$,
D. Atri$^{3}$ and
B. E. Wood$^{4}$
\\
$^{1}$School of Physics, Trinity College Dublin, The University of Dublin, Dublin 2, Ireland
\\ $^{2}$Leiden Observatory, Leiden University, PO Box 9513, NL-2300 RA Leiden, The Netherlands
\\ $^{3}$Center for Space Science, New York University Abu Dhabi, PO Box 129188, Saadiyat Island, Abu Dhabi, UAE
\\ $^{4}$Naval Research Laboratory, Space Science Division, Washington, DC 20375, USA}

\date{Accepted XXX. Received YYY; in original form ZZZ}

\pubyear{2021}

\begin{document}
\label{firstpage}
\pagerange{\pageref{firstpage}--\pageref{lastpage}}
\maketitle

\begin{abstract}
Quantifying the flux of cosmic rays reaching exoplanets around M dwarfs is essential to understand their possible effects on exoplanet habitability. Here, we investigate the propagation of Galactic cosmic rays as they travel through the stellar winds (astrospheres) of five nearby M dwarfs, namely: GJ~15A, GJ~273, GJ~338B, GJ~411 and GJ~887. Our selected stars each have 1 or 2 detected exoplanets and they all have wind mass-loss rates constrained by Lyman-$\alpha$ observations. Our simulations use a combined 1D magnetohydrodynamic (MHD) Alfvén-wave-driven stellar wind model and 1D cosmic ray transport model. We find that GJ~411 and GJ~887 have Galactic cosmic rays fluxes comparable with Earth’s at their habitable zones. On the other hand, GJ~15A, GJ~273 and GJ~338B receive a lower Galactic cosmic ray flux in their habitable zones. All exoplanets in our sample, with exception of GJ~15A\,c and GJ~411\,c, have a significantly lower flux of Galactic cosmic rays than values observed at the Earth because they orbit closer-in. The fluxes found here can be further used for chemical modelling of planetary atmospheres. Finally, we calculate the radiation dose at the surface of the habitable-zone planet GJ~273\,b, assuming it has an Earth-like atmosphere. This planet receives up to 209 times less 15\,MeV energy cosmic ray fluxes than values observed at Earth. However, for high-energy cosmic rays ($\sim$\,GeV), the difference in flux is only 2.3 times smaller, which contributes to GJ~273\,b receiving a significant surface radiation dose of 0.13\,mSv/yr (40\% of the annual dose on Earth's surface). 
\end{abstract}




\section{Introduction}
With more than 4,300 exoplanets discovered and confirmed (as of 30 September 2021 on NASA's exoplanet archive\footnote{\url{https://exoplanetarchive.ipac.caltech.edu}}) in the last few decades, there is a lot of interest in discovering/determining if any of these exoplanets are habitable. There are many factors affecting the habitability of an exoplanet \citep[see, e.g.,][]{Meadows2018}. One key factor is the presence of liquid water on the exoplanet surface. Many factors can influence the presence of surface liquid water on a planet, such as the planet and stellar system properties. The habitable zone is defined as the orbital distances from a star where liquid water can exist on a planet's surface \citep{Kasting1993,Selsis2007}. This region is close-in for M dwarfs and further out for F and G dwarfs. Although being in the habitable zone does not necessarily mean that a planet is habitable, this is the first condition thought to be important for habitability.

Currently M dwarfs are the main targets in the search for potential habitable exoplanets. This is because M dwarfs are small, they have low brightness, and consequently, they have a close-in habitable zone. This combination, due to current observational capabilities, makes M dwarfs the best candidates to observe exoplanets in the habitable zone. Additionally, they constitute the majority of stars in our Galaxy \citep{Henry2006, Winters2015, Henry2018}. However, the stellar environment of M dwarfs can be very harmful for close-in exoplanets. M dwarfs can produce strong magnetic fields \citep{Morin2010, Shulyak2019}, and compared with solar-mass stars, they stay magnetically active for a longer duration of their lives \citep{West2004, Scalo2007, West2015, Guinan2016}. Exoplanet habitability can be affected by strong stellar activity \citep{Khodachenko2007, Vida2017, Tilley2019}. This is because, strong stellar activity leads to stronger stellar winds \citep{Vidotto2014}, generates more stellar energetic particles \citep{Griebmeier2005}, stronger flares \citep{Vida2017, Tilley2019} and coronal mass ejections \citep{Lammer2007, Khodachenko2007}, all of which can affect planetary atmospheres and thus their potential to generate life.

In particular, the stellar wind plays an important role in stellar evolution \citep{Johnstone2015, Matt2015} and interacts with planets \citep{Vidotto2011, Vidotto2013, Vidotto2020}. For this reason, it is important to understand the properties of stellar winds. M dwarfs, however, have rarefied coronal winds analogous to the solar wind, which makes it difficult to observe them. Fortunately, novel techniques have been developed to characterise the winds of low-mass stars \citep[see review by][]{Vidotto2021}  providing some constraints on the wind properties, such as, the stellar wind mass-loss rate. These methods include Lyman-$\alpha$ absorption observations \citep[][and references therein]{Wood2004, Wood2021}, X-ray emission \citep{Wargelin2001, Wargelin2002}, radio emission \citep{Panagia1975, Lim1996, Fichtinger2017, VidottoD2017}, exoplanet atmospheric escape \citep{Vidotto2017, Kislyakova2019} and slingshot prominences \citep{Jardine2019}. Despite all of these methods being able to provide constraints for a few tens of low-mass stars there are still unknowns related to the stellar wind properties. 

In our work, we study the properties of M dwarf winds using numerical simulations and observational constraints from Lyman-$\alpha$ measurements. Here, we assume the M dwarf winds are heated and accelerated by the dissipation of Alfv\'{e}n waves. We use an Alfv\'{e}n-wave-driven stellar wind model to understand the stellar wind and calculate its properties, such as, the velocity profile and mass-loss rate \citep{Mesquita2020}. 

One particular aspect of interest in our work is that stellar winds affect the propagation of cosmic rays, which are an important factor that could affect exoplanet habitability. On the one hand, cosmic rays may have been important for the origin of life on Earth and for other exoplanets \citep{Airapetian2016, Atri2016}, as they can drive the production of prebiotic molecules \citep{Rimmer2014, Airapetian2016, Barth2020}. In contrast, for developed life-forms cosmic rays can damage DNA in cells \citep{Sridharan2016} and cause cellular mutation \citep{Dartnell2011}. Thus, large cosmic ray fluxes are harmful for life as we know it \citep{Shea2000}. A way to quantify the impact of cosmic rays on life-forms is by calculating the radiation dose on the surface of a planet \citep{Atri2020, Atri2020-2}. In addition to this, cosmic rays can also affect cloud coverage which could affect Earth's climate \citep{Svensmark1997, Shaviv2002, Shaviv2003, Kirkby2011, Svensmark2017}. All of these aspects about cosmic rays makes it important to investigate their effects on exoplanets.

There are two populations of cosmic rays: Galactic cosmic rays and stellar cosmic rays generated by the host star. Many works have focused on the impact of stellar cosmic rays fluxes at the Earth \citep{Rodgers2021}, on M dwarfs \citep{Fraschetti2019}, on young T-tauri systems \citep{Rab2017, Rodgers2017}, on exoplanets' magnetospheres and atmospheres \citep{Segura2010, Grenfell2012, Tabataba2016, Scheucher2020} and also in the context of star-forming regions \citep[see review by][]{Padovani2020}. In this work, we only consider the effects of Galactic cosmic rays originating from our own Galaxy.

Many studies have investigated the Galactic cosmic ray fluxes at Earth focusing on different ages of the solar system \citep{Scherer2002, Scherer2008, Muller2006, Svensmark2006, Cohen2012, Rodgers2020} to understand their possible effects on Earth. More recently, \citet{Rodgers2021-2} studied the Galactic cosmic rays fluxes for a well-constrained sample of five Sun-like stars with magnetic field measurements and Lyman-$\alpha$ observations. Some works \citep{Sadovski2018, Herbst2020, Mesquita2021} have also analysed the Galactic cosmic ray fluxes at exoplanets orbiting M dwarfs. 

\begin{figure}
	\includegraphics[width=\columnwidth]{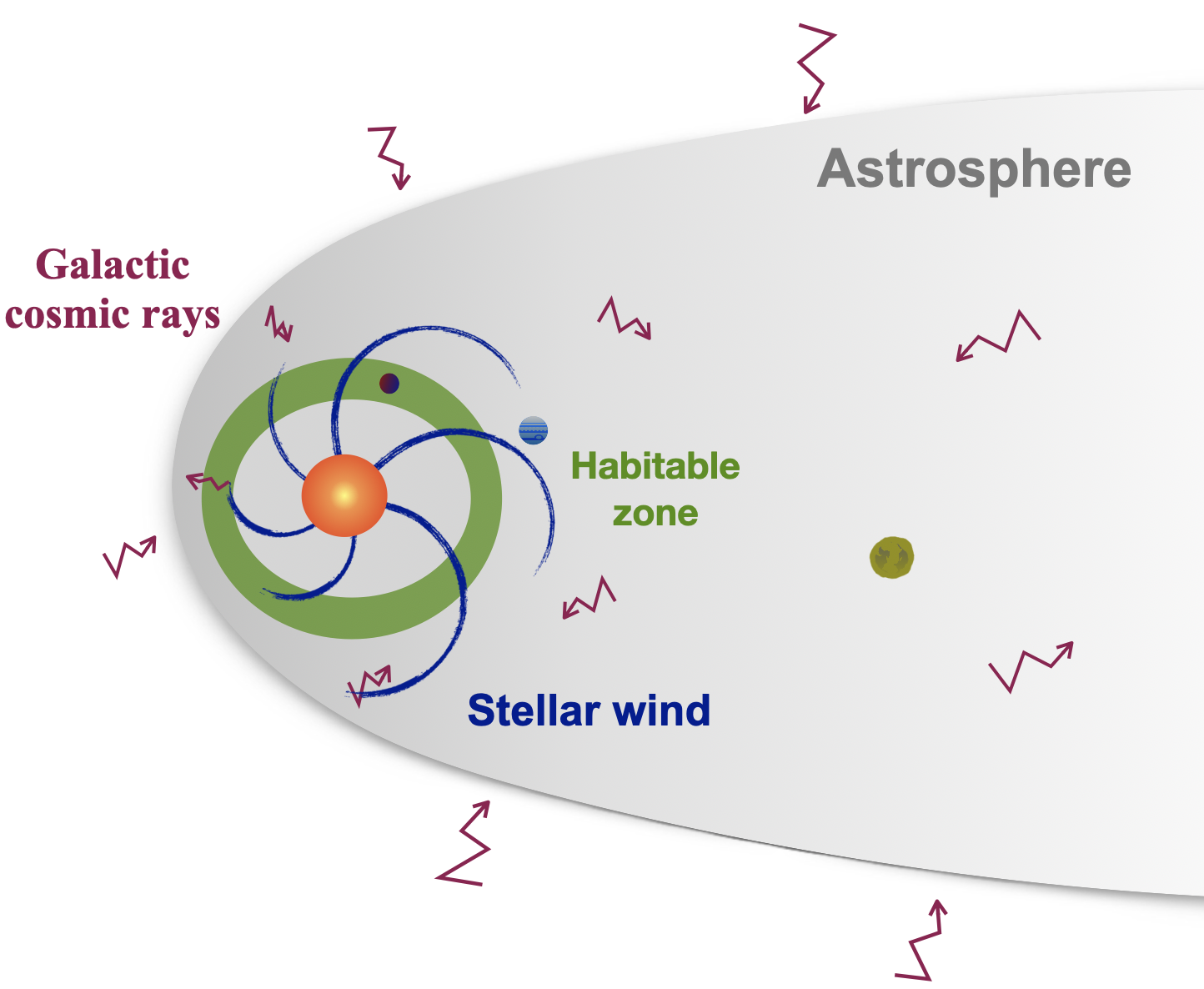}
    \caption{Schematic of the stellar system environment studied in this work. The wind generated by the star (large yellow circle) interacts with the interstellar medium (ISM) and creates a ``bubble'' region called the astrosphere (grey region). The Galactic cosmic rays propagate inside the astrosphere and can interact with the host planets (small circles).}
    \label{fig:sketch}
\end{figure}

The interaction between the stellar wind and the interstellar wind forms a bubble shaped region around the star which is dominated by its stellar wind (see \Cref{fig:sketch}). This region is called the astrosphere, analogous to the Sun's heliosphere. When studying the propagation of Galactic cosmic rays, the astrosphere becomes especially relevant -- outside this region, the cosmic ray flux has its background level. The astrosphere acts as a barrier to the cosmic rays due to the presence of a magnetised stellar wind. Within the astrosphere, the flux of Galactic cosmic rays is modulated/suppressed in a energy-dependent way by the stellar wind.

In this paper, we investigate the propagation of Galactic cosmic rays through a sample of five M dwarf astrospheres. We focus on M dwarfs that host known exoplanets and have mass-loss rates constrained by Lyman-$\alpha$ observations recently presented in \citet{Wood2021}. We perform 1D MHD simulations, using an Alfvén-wave-driven stellar wind model to derive the  properties of the winds of the M dwarfs \citep{Mesquita2020}. We use a 1D model of cosmic ray transport \citep[from][]{Rodgers2020} to calculate the spectrum of Galactic cosmic rays within M dwarf astrospheres. This paper is organised as follows: in \Cref{sec:sample} we describe the stars in our sample. The stellar wind model and the astrospheric sizes are presented in \Cref{sec:sw}. The transport model for the cosmic rays is described in \Cref{sec:cr}. Our results on the Galactic cosmic ray fluxes in the habitable zone and at the exoplanets' orbits, together with the radiation dose at GJ~273\,b's surface are shown in \Cref{sec:results}, followed by our discussion and conclusions in \Cref{sec:conc}. 

\section{The stars in our sample}
\label{sec:sample}
We select stars with well-constrained stellar wind ram pressure values
from recent Lyman-$\alpha$ observations presented in \citet{Wood2021}. Our main targets are M dwarfs because they have been identified as prime targets in the search for bio-signatures from exoplanets \citep{Meadowsp2018}. Additionally, our sample only contains M dwarfs with at least one known orbiting exoplanet. Using these criteria we select five M dwarfs, namely: GJ~15A, GJ~273, GJ~338B, GJ~411 and GJ~887. All of these stars are in close proximity to the solar system, within 7\,pc. \Cref{tab:stars} presents the stellar properties relevant for our stellar wind simulations, such as radius, mass, distance, rotation period and X-ray luminosity. \Cref{tab:planets} summarises the list of known exoplanets in our sample of stars and their properties.

\begin{table*}
 \caption{Sample of stars studied in our work. The parameters are from \citet{Wood2021} unless explicitly stated otherwise. The columns are, respectively, star ID, stellar mass, radius, stellar spectral type, rotation period, X-ray luminosity, ISM velocity as seen by the star, mass-loss rate, wind ram pressure, ISM ram pressure and habitable zone boundaries. The last two columns are quantities calculated in our work.}
 \begin{tabular}{lccccccccccc}
  \hline
  Star & Mass & Radius &  Spectral & $P_{\text{rot}}$ & $\log L_X$ & $v_\text{ISM}$ & $\dot{M}_{\rm wood}$ & $P_{\rm ram,\ wood}(300\,R_\star)$ & $P_{\rm ISM}$ & Habitable zone\\
   & [$M_{\sun}$] & [R$_{\sun}$] & type & [days] & [erg\,s$^{-1}$] & [km~s$^{-1}$] & [$M_{\odot}~\text{yr}^{-1}$] & [dyn~cm$^{-2}$] & [dyn~cm$^{-2}$] & [au]\\
  \hline
  GJ~15A & 0.38 $^a$ & 0.38 $^a$ & M2 V & 44 $^b$ & 27.37 & 28 & $2 \times 10^{-13}$ & $6.4 \times 10^{-7}$ & $3.2 \times 10^{-12}$ & 0.12 -- 0.31\\
  GJ~273 & 0.29 $^c$ & 0.293 $^c$ & M3.5 V & 99 $^c$ & 26.54 & 75 & $< 4 \times 10^{-15}$ & $<2.2 \times 10^{-8}$ & $2.4 \times 10^{-11}$ & 0.08 -- 0.20\\
  GJ~338B & 0.64 $^d$ & 0.58 $^d$ & M0 V & 16.6 $^d$ & 27.92 & 29 & $1 \times 10^{-14}$ & $1.4 \times 10^{-8}$ & $3.4 \times 10^{-12}$ & 0.23 -- 0.58\\
  GJ~411 & 0.386 $^e$ & 0.389 $^e$ & M2 V & 56.2 $^e$ & 26.89 & 110 & $<2 \times 10^{-15}$ & $<6.1 \times 10^{-9}$ & $4.7 \times 10^{-11}$ & 0.12 -- 0.31\\
  GJ~887 & 0.489 $^f$ & 0.471 $^f$ & M2 V & $>200\ ^f$ & 27.03 & 85 & $1 \times 10^{-14}$ & $2.1 \times 10^{-8}$ & $2.9 \times 10^{-11}$ & 0.16 -- 0.40\\
  \hline
 \end{tabular}
 \label{tab:stars}
 \newline
 $^a$ \citet{Pinamonti2018}; $^b$ \citet{Howard2014}; $^c$ \citet{Astudillo2017}; $^d$ \citet{Gonzalez2020}; $^e$ \citet{Diaz2019}; $^f$ \citet{Jeffers2020}.
\end{table*}

\begin{table}
 \caption{Properties of the known exoplanets in our sample of stars. The columns are, respectively, planet name, semi-major axis, mass, orbital period and references for the properties.}
 \begin{tabular}{lcccc}
  \hline
  Planet & $a$ & $M_p$ & $P$ & References\\
   & [au] & [$M_\oplus$] & [days] \\
     \hline
   GJ~15A\,b & 0.072 & 3.03 & 11.44 & 1\\
   GJ~15A\,c & 5.4 & 36 & 7600 & 1\\
   GJ~273\,b & 0.091 & 2.89 & 18.65 & 2\\
   GJ~273\,c & 0.036 & 1.18 & 4.72 & 2\\
   GJ~338B\,b & 0.141 & 10.27 & 24.45 & 3\\
   GJ~411\,b & 0.079 & 2.69 & 12.95 & 4\\
   GJ~411\,c & 3.10 & 18.1 & 3190 & 5\\
   GJ~887\,b & 0.068 & 4.2 & 9.26 & 6\\
   GJ~887\,c & 0.12 & 7.6 & 21.79 & 6\\
  \hline
 \end{tabular}
 \label{tab:planets}
 \newline
    (1) \citet{Pinamonti2018}; (2) \citet{Astudillo2017}; (3) \citet{Gonzalez2020}; (4) \citet{Stock2020}; (5) \citet{Rosenthal2021}; (6) \citet{Jeffers2020}.
\end{table}

GJ~15A and GJ~338B are wide-orbit binary systems with a second M dwarf. Their orbital separations are  $146\,\text{au}$ \citep{Pinamonti2018} and $110\,\text{au}$ \citep{Gonzalez2020}, respectively. However, this is still close enough for each pair to reside within a common astrosphere, meaning the measured wind strength represents the combined wind of both stars \citep{Wood2021}. For the purposes of our work, when simulating the wind and the cosmic ray transport we assume that the wind is being produced by GJ~15A and GJ~338B, because they are the companions with planets.

\section{The stellar wind environment}
\label{sec:sw}
\subsection{The Alfvén-wave-driven wind simulation}
\label{sec:AWDW}
To investigate the stellar wind properties for each star in our sample we use 1D time-independent MHD simulations. Similar to the solar wind, we assume the winds of M dwarfs are heated by magnetic processes. Here, the stellar wind is heated and accelerated by the dissipation of Alfvén waves. These waves are generated due to perturbations of the magnetic field lines at the base of the wind. The wind is launched at the base of the chromosphere and the grid extends until $300\,R_{\star}$. Note that all the stars in our sample have reached their terminal velocities by this distance. The model used here is based on the model presented in \citet{Mesquita2020} \citep[see also][]{Vidotto2010}.

The time-independent MHD equations are given as follows:
\begin{equation}
    \frac{d}{dr}\left(\rho u r^2\right) = 0,
    \label{eq:mass-con}
\end{equation}
\begin{equation}
    u\frac{du}{dr}=-\frac{GM_{\star}}{r^2}-\frac{1}{\rho}\frac{dP}{dr}-\frac{1}{2\rho}\frac{d\epsilon}{dr},
    \label{eq:momentum}
\end{equation}
\begin{equation}
    \rho u\frac{d}{dr}\left(\frac{u^2}{2}+\frac{5}{2}\frac{k_{B}T}{m}-\frac{GM_{\star}}{r}\right)+\rho u \frac{d}{dr}\left(\frac{F_{c}}{\rho u}\right)+\frac{u}{2}\frac{d\epsilon}{dr}=Q-P_{r},
    \label{eq:energy}
\end{equation}
where $r$ is the radial coordinate, $\rho$ the wind mass density, $u$ the wind velocity, $G$ the gravitational constant, $M_{\star}$ the stellar mass, $P=\rho k_B T/m$ the gas pressure, $m$ the average mass of the wind particles, $T$ the wind temperature, $\epsilon$ the energy density of the Alfv\'{e}n waves, $F_c$ the thermal conduction, $Q$ the heating term and $P_r$ is the radiative cooling term. \Cref{eq:mass-con,eq:momentum,eq:energy} are the mass, momentum and energy conservation equations, respectively. The terms on the right-hand side of \Cref{eq:momentum} are the gravitational, thermal and mechanical forces, respectively. The terms inside the first parentheses on the left-hand side of \Cref{eq:energy} are the kinetic, enthalpy and gravitational energies per unit mass which are associated with the wind energy. The second term is the conductive energy and the third is the wave energy.

The parameters required in our model are the magnetic field strength and geometry, damping type and length, stellar wind density, temperature and magnetic field perturbation intensity. All the input parameters in our simulations are defined at the base of the chromosphere. We refer the reader to \citet{Mesquita2020}, who investigated how these parameters influence the properties of the stellar wind, such as density, velocity, temperature and mass-loss rate. Here, we use a fully radial magnetic field line configuration. We use the same temperature of $10^4$\,K at the base of the chromosphere for all of the stars. We adopt a non-linear damping mechanism in which the amplitude of the MHD waves decreases with the quadratic amplitude of the fluctuations in the wave velocity, using the approach by \citet{Jatenco1989}. We adopt an initial damping length of $0.1\,R_{\star}$ \citep{Mesquita2020}. The non-linear damping mechanism for the waves has been used in solar wind models \citep[e.g.,][]{Suzuki2005,Suzuki2013}. The magnetic field perturbations are taken at the base of the wind as $0.1\,B_0$, where $B_0$ is the magnetic field strength at the base of the wind. The magnetic field perturbation at the base of the wind is a free parameter in our simulation that can affect the properties of the stellar wind \citep[see discussion in][]{Mesquita2020}. The amplitude of the magnetic field fluctuations is connected with the amplitude of velocity fluctuations by energy equipartition \citep[see Equation 5 in][]{Mesquita2020}. \citet{Holst2014} used the velocity fluctuations value of 15\,km/s at the chromosphere for a 3D MHD solar simulation. This value agrees with Hinode observations of the 15\,km/s turbulent velocities for the solar wind \citep{Depontieu2007}. In our simulations, the velocity fluctuations at the base of the wind vary from 9--23\,km/s (using the magnetic field perturbations at the base of the stellar wind equal to $0.1\,B_0$) which is in good agreement with the works mentioned for the solar wind. We use observations to constrain the magnetic field strength and the density at the base of the stellar wind in our model. The approach used here is explained below.

\subsubsection{Stellar surface magnetic field}
In order to constrain the input magnetic field strength at the base of the chromosphere we use the observed correlation between X-ray luminosity and large scale magnetic flux, $L_X \approx 10^{-13.7} \Phi_{V}^{1.80\pm 0.20}$, from \citet{Vidotto2014}. We use the X-ray luminosity given in \Cref{tab:stars} to infer the large scale magnetic flux using this relation. Then, we use the relation $\Phi_V=4\pi R_{\star}^2 B_0$ to get the average large scale magnetic field strength. This value of $B_0$ is used as the input magnetic field strength for the stellar wind simulations. The value used for each star is presented in \Cref{tab:stars2}. The stars in our sample are old and not very active, which explains the relatively low magnetic field strength values (3.1--7.5\,G) found in our work \citep[consistent with recent spectropolarimetric results from][]{Moutou2017}. More active M dwarfs can have kilo Gauss magnetic field strengths \citep[e.g. review by][]{Morin2012}. 

\begin{table*}
 \caption{Stellar wind properties. The columns are, respectively, the star ID, the magnetic field strength and density at the base of the wind (input model parameters), terminal velocity, mass-loss rate, stellar wind ram pressure, Alfvén radius and astrosphere size (output model parameters).}
 \begin{tabular}{lcccccccc}
  \hline
  Star & $B_0$ & $\rho_0$ & $u_\infty$ & $\dot{M}$ & $P_{\rm ram}(300\,R_\star)$ & $R_{\rm A}$ & $R_{\rm ast}$\\
   & [G] & [g~cm$^{-3}$] & [km/s] & [$M_{\odot}~\text{yr}^{-1}$] & [dyn~cm$^{-2}$] & [au] & [au]\\
  \hline
  GJ~15A & 7.5 & $2.6 \times 10^{-14}$ & 1100 & $6.8 \times 10^{-14}$ & $6.2 \times 10^{-7}$ & 0.02 & 240\\
  GJ~273 & 4.3 & $9.8 \times 10^{-15}$ & 880 & $<1.8 \times 10^{-15}$ & $<2.2 \times 10^{-8}$ & 0.06 & 12\\
  GJ~338B & 6.0 & $5.5 \times 10^{-15}$ & 940 & $1.1 \times 10^{-14}$ & $3.6 \times 10^{-8}$ & 0.06 & 52\\
  GJ~411 & 3.5 & $2.4 \times 10^{-15}$ & 1400 & $<6.1 \times 10^{-16}$ & $<6.4 \times 10^{-9}$ & 0.08 & 6\\
  GJ~887 & 3.1 & $9.6 \times 10^{-15}$ & 850 & $4.8 \times 10^{-15}$ & $2.1 \times 10^{-8}$ & 0.05 & 18\\
  \hline
 \end{tabular}
 \label{tab:stars2}
\end{table*}

\subsubsection{The base density of the stellar wind}
One successful technique used to detect stellar winds of low-mass stars is the detection of Lyman-$\alpha$ absorption \citep[][and references therein]{Wood2021}. Lyman-$\alpha$ absorption is generated when  stellar photons travel through the stellar astrosphere, the ISM and the heliosphere and is detected in UV spectra. With the detection of astrospheric absorption it is possible to determine the ram pressure of the stellar wind, $P_{\text{ram}}$. By knowing the stellar wind asymptotic velocity, $u_\infty$, it is possible to determine the mass-loss rate, $\dot{M}$, using the relation
\begin{equation}
    \dot{M} = \frac{4\pi R_\star^2 P_{\text{ram}}}{u_\infty}.
    \label{eq:wood}
\end{equation}
On the other hand, if astrospheric absorption is not detected only an upper limit for $P_{\text{ram}}$ is obtained. In both cases, the Lyman-$\alpha$ observations are essential to constrain the density at the base of the chromosphere that we use in our stellar wind model.

For that, we vary the value of the density at the base of the wind to reproduce the stellar wind ram pressure inferred from the Lyman-$\alpha$ observations. We match the stellar wind ram pressure, rather than the mass-loss rates presented in  \citet{Wood2021}, because  \citet{Wood2021} assumes the Sun's terminal wind velocity of 400\,km/s for all the stars. In our simulations, the values we find for $u_\infty$ are larger. The ram pressures derived from the observationally-inferred mass-loss rates and $u_\infty=400$~km/s are shown in column ten of \Cref{tab:stars} ($P_{\rm ram,\ wood}$) -- these values are calculated at a reference radius of $300~R_\star$. For comparison, we show the ram pressure values that our models produce in \Cref{tab:stars2} (calculated at the same reference radius). As it can be seen, our values are reasonably similar (i.e., within a factor of 2.6) from the observationally-derived ones. In terms of mass-loss rates, our approach gives lower values than those found in \citet{Wood2021} (see $\dot{M}_{\rm wood}$ in \Cref{tab:stars}) due to our higher stellar wind terminal velocities. The value of the base density, $\rho_0$, used for each simulation is shown in \Cref{tab:stars2}.

\subsection{Stellar wind density, velocity and magnetic field profiles}
The stellar wind properties, such as the magnetic field and velocity profiles, are needed to calculate the Galactic cosmic ray flux reaching an exoplanet. From our Alfvén-wave-driven wind simulations we have the stellar wind velocity profile. However, only the radial component of the magnetic field, $B_r$, is calculated from our 1D MHD simulations. Thus, to determine the azimuthal component of the magnetic field, $B_\phi$, we use the Parker spiral relation \citep{Parker1958}:
\begin{equation}
    \frac{B_\phi}{B_r}=\frac{u_\phi -r\Omega}{u_r},
    \label{eq:spiral}
\end{equation}
where $u_\phi$ is the azimuthal velocity component, $\Omega=2\pi/P_{\text{rot}}$ is the angular speed of the star and $P_{\text{rot}}$ is the stellar rotation period. $B_\phi$ only dominates at large distances and at these distances, $u_\phi\ll r\Omega$. Thus, \Cref{eq:spiral} can be expressed as 
\begin{equation}
    |B_\phi|\simeq \frac{r\Omega}{u_r}B_r.
    \label{eq:bphi}
\end{equation}
The total magnetic field strength is $B= \sqrt{B_r^2+B_\phi^2}$, where $B_\phi$ is given by \Cref{eq:bphi} for distances beyond the Alfvén radius, $r>R_A$. The Alfvén radius is defined as the distance where the wind has reached the Alfvén velocity which is given by $v_A= B/\sqrt{4\pi \rho}$.

The outer boundary of the Alfvén-wave-driven wind simulations is at 300\,$R_\star$, where the wind has already reached its terminal velocity. However, the astrosphere extends further out and we extrapolate the quantities $u_r$, $B_r$ and $B_\phi$ to take into account the profiles for the whole astrosphere. Since the velocity reaches its asymptotic value by 300\,$R_\star$, beyond this distance it has a constant profile. The radial magnetic field component continues to fall with $r^2$ and the azimuthal component with $r$, generating the Parker spiral \citep{Parker1958}.

\begin{figure*}
\includegraphics[width=\textwidth]{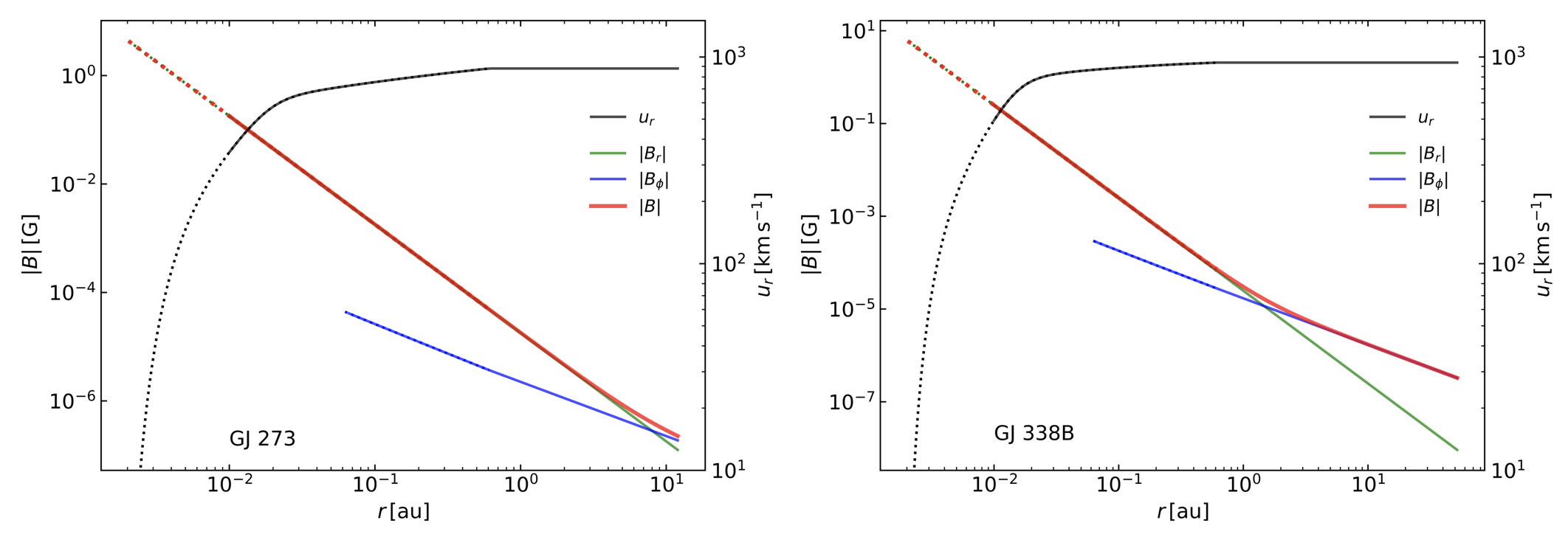}
    \caption{Stellar wind properties, such as, velocity (black curves) and magnetic field (total magnetic field red curves, radial magnetic field strength green curves and azimuthal magnetic field blue curves) profiles of left: GJ~273 and right: GJ~338B. The dotted lines are the results obtained from the stellar wind model and the solid lines are the profiles used as input in the cosmic ray model.}
    \label{fig:wind_results}
\end{figure*}

\Cref{fig:wind_results} shows a summary of the stellar wind properties for GJ~273 (left) and GJ~338B (right) to show the different contributions of the magnetic field components. In both panels, the black curve is the velocity profile, the red curve is the total magnetic field, the green curve is the radial magnetic field strength and the blue curve is the azimuthal magnetic field strength. The dotted curves are the results from the Alfvén-wave-driven wind simulations and the solid curves are the profiles used as an input for the cosmic ray simulations. Two planets in our simulations, GJ~273\,c and GJ~411\,b, lie within the Alfv\'{e}n radius, in a sub-Alfv\'{e}nic region.

\subsection{The size of the M dwarf astrospheres}
The outer boundary of the cosmic ray simulation is set to be the astrospheric radius, which varies for the stars in our sample. The size of the astrosphere can be calculated by finding the balance between the stellar wind ram pressure and the ISM ram pressure. The wind ram pressure is given by:
\begin{equation}
    P_{\text{ram}}=\rho u^2.
    \label{eq:ram-wind}
\end{equation}
The ISM ram pressure is given by:
\begin{equation}
    P_{\rm ISM}=m_p n_{\rm ISM}\nu_{\rm ISM}^2,
    \label{eq:ram-ism}
\end{equation}
where $m_p$ is the proton mass, $n_{\text{ISM}}$ is the total ISM number density of hydrogen and $\nu_{\text{ISM}}$ is the ISM velocity as seen by the star. 

The ISM ram pressure was calculated using $\nu_{\text{ISM}}$ for each star as given in \Cref{tab:stars}. The total ISM number density of hydrogen is given by the sum of neutral hydrogen number density and the ionised hydrogen number density ($n_{\rm ISM} = n_{\rm n}+n_{\rm i}$). We assume that the ISM is partially ionised and we use the values from \citet{Wood2000} that successfully reproduces heliospheric absorption, e.g., Model 10 of \citet{Wood2000}, which are $n_{\rm n}=0.14\,\text{cm}^{-3}$ and $n_{\rm i}=0.1\,\text{cm}^{-3}$, giving $n_{\rm ISM}=0.24\,\text{cm}^{-3}$ for all stars in our sample. Using \Cref{eq:ram-ism} we calculate $P_{\rm ISM}$ for each star in our sample, the results are shown in \Cref{tab:stars}. 

The astrospheric size is calculated as:
\begin{equation}
    R_\text{ast}=\sqrt{\frac{P_{\text{ram}}(R)}{P_{\text{ISM}}}}R,
    \label{eq:astro}
\end{equation}
where $R$ is a given reference distance where the wind has reached its terminal velocity. The stellar wind ram pressure calculated at $R=300\,R_\star$ is shown in \Cref{tab:stars2}. In general, M dwarfs tend to be an older stellar population due to their long lifetimes \citep[see review by][]{Shields2016}, and consequently, they will on average have higher $\nu_{\text{ISM}}$ values because as stars pick up more gravitational perturbations with time they acquire larger deviations from the Local Standard of Rest. For this reason, M dwarfs tend to have small astrospheres. 

Using \Cref{eq:astro} we calculate the astrospheric size of each star in our sample, the values are shown in \Cref{tab:stars2}. The size of the astrospheres in our sample vary from 6 to 240\,au. GJ~411 has the smallest astrospheric size and is a very compact system. In constrast, GJ~15A has a very large astrosphere being almost as twice as large as the present-day heliosphere \citep[$\sim 122$\,au,][]{Stone2013, Stone2019}. Note that for GJ~338B, we obtain a larger astrosphere size of 83\,au instead of the 52\,au using $P_{\text{ram,\ wood}}$. This is because our value for $P_{\text{ram}}$ is around 2.6 times larger than the value $P_{\rm ram,\ wood}$ \footnote{We found that this difference in the astrospheric size of GJ~338B does not strongly affect the Galactic cosmic ray fluxes for the system. A similar situation was also observed in the GJ~436 stellar system \citep[discussed in][]{Mesquita2021}.}.

\section{Cosmic ray transport}
\label{sec:cr}
As Galactic cosmic rays propagate through a magnetised stellar wind they suffer global variations in their intensity and energy which is known as the modulation (or suppression) of cosmic rays. The modulation of cosmic rays can be obtained by solving the diffusion-advection transport equation of \citet{Parker1965}. The model we use was presented in \citet{Rodgers2020} and was previously applied to the M dwarf, GJ~436 \citep{Mesquita2021}. We numerically solve the time-dependent transport equation 
\begin{equation}
    \frac{\partial f}{\partial t}=\nabla\cdot(\kappa \nabla f) - u\cdot(\nabla f)+\frac{1}{3}(\nabla\cdot u)\frac{\partial f}{\partial \ln p},
    \label{eq:transport}
\end{equation}
where $f(r,p,t)$ is the cosmic ray space phase density, $r$ is the radial distance, $p$ is the cosmic ray momentum and $u$ is the stellar wind velocity. The first term on the right-hand side of \Cref{eq:transport} represents the diffusion of the cosmic rays which depends on the diffusion coefficient, $\kappa (r,p)$. The second term is the advection of the cosmic rays which depends on the stellar wind velocity, $u$, and acts against the inward diffusion of the cosmic rays. The third term is the adiabatic losses due to the stellar wind expansion \footnote{Here, we do not take into account ionisation losses because the cosmic rays do not have a lot of material to cross when travelling in the astrosphere of the M dwarfs in our sample. This is because the column density of stellar wind material is small. For instance, for GJ~436 we calculated, using the stellar wind density, that the column density of stellar wind material is $2.8\times 10^{-7}\,\text{g~cm}^{-2}$.}.

Stellar wind magnetic field (via the diffusion coefficient) and velocity profiles are key ingredients for the modulation of Galactic cosmic rays, as given by \Cref{eq:transport}. The stellar wind density profile is relevant to define the size of the astrosphere. However, it does not lead to any significant attenuation of the cosmic rays, as the stellar wind wind density is very low.

Our spatial and momentum grids are logarithmically spaced with 60 grid zones each. The spatial inner boundary is 0.01\,au and the outer boundary is set as the astrospheric size of each star in our sample (see \Cref{tab:stars2}).  The momentum range that we consider for our simulations is $p_{\rm min}=0.15$\,GeV$/c$ and $p_{\rm max}=100$\,GeV$/c$. The upper limit for the momentum was selected because particles with energies above this limit are very infrequent and are not relevant in the context of planetary atmosphere chemistry \citep{Rimmer2013}. The lower limit was selected because particles with low energies ($\lesssim 290$\,MeV, i.e. the pion threshold energy) do not reach the planetary surface \citep{Atri2017} and do not contribute to the radiation dose calculated there  (this will be calculated in \Cref{sec:radiation}). However, the low-energy particles are important in the context of planetary atmospheres because they deposit all of their energy there \citep{Rodgers2020} and could be included in future chemical modelling studies.

\subsection{Diffusion coefficient}
As Galactic cosmic rays penetrate a stellar astrosphere their intensity is reduced due the presence of a magnetised stellar wind. The diffusion of the cosmic rays depends on the turbulence level of the magnetic field. The presence of magnetic field irregularities makes the cosmic ray undergo a random walk in the system. The diffusion coefficient of the cosmic rays, from quasi-linear theory \citep{Jokipii1966, Schlickeiser1989}, can be expressed as
\begin{equation}
  \frac{\kappa(r,p)}{\beta c}=\eta_0\left(\frac{p}{p_0}\right)^{1-\gamma}r_\text{L},
\end{equation}
where $\beta=v/c$ is the ratio between the particle velocity and the speed of light, $p_0=3\,$GeV$/c$, $r_\text{L}=p/eB(r)$ is the Larmor radius of the protons. $\eta_0$ depends on the level of turbulence in the magnetic field. We adopt $\eta_0=1$ which represents the maximum level of turbulence for the magnetic field. $\gamma$ determines how the diffusion coefficient varies with momentum. We adopt $\gamma=1$, which corresponds to Bohm diffusion. This value is commonly used in other works \citep{Svensmark2006, Cohen2012, Rodgers2020} and is in good agreement with observations at Earth. There are different prescriptions for $\gamma$, such as, the Kolmogorov-type turbulence \citep[$\gamma=5/3$, as in][]{Herbst2020} and magnetohydrodynamic-driven turbulence ($\gamma=3/2$). The type of turbulence for M dwarf systems is currently unknown. The turbulence type can affect the cosmic ray spectrum for all energies. For instance, for Kolmogorov-type turbulence normalised at 1\,GeV/c, \citet{Mesquita2021} found that cosmic rays with momentum <1\,GeV/c are less modulated when compared with Bohm-type turbulence. On the other hand, cosmic rays with momentum >1\,GeV/c are more modulated. 

\subsection{Local Interstellar Spectrum (LIS)}
In the ISM there is a ``sea'' of Galactic cosmic rays, unaffected by the presence of the magnetised stellar wind. This value sets the background flux of Galactic cosmic rays that can penetrate the stellar systems. 

The unaffected background spectrum of Galactic comic rays was observed by {\it Voyager 1}, after it crossed the heliopause \citep{Stone2013, Cummings2016}. In our simulations, we use the fit to the local interstellar spectrum (LIS) from \citet{Vos2015} to describe the unaffected spectrum of Galactic cosmic rays that can be injected in our system. Using {\it Voyager 1} observations, \citet{Vos2015} developed a model fit to describe the LIS:
\begin{equation}
    j_{\text{LIS}}(T)=2.70\frac{T^{1.12}}{\beta^2}\left(\frac{T+0.67}{1.67} \right)^{-3.93}\,\text{m}^{-2}\text{s}^{-1}\text{sr}^{-1}\text{MeV}^{-1},
    \label{eq:jlis}
\end{equation}
where $j$ is the differential intensity of cosmic rays and $T$ is the kinetic energy of the cosmic rays in GeV. In our simulations, the LIS is considered constant as a function of time. Note, the differential intensity can be expressed in terms of the phase space density, $f$ in \Cref{eq:transport}, as $j(T)=p^2f(p)$.

$\gamma$-ray observations, which trace $\sim 10-10^{4}$\,GeV cosmic rays, within 1 kpc in the local Galaxy inferred the Galactic cosmic ray spectrum to be in a good agreement with {\it Voyager} measurements in the local ISM \citep{Neronov2017}. However, ionisation rates inferred from observations of diffuse molecular clouds \citep[see discussion in][]{Recchia2019, Padovani2020}, which trace lower energy cosmic rays ($\lesssim$\,GeV), indicate that there are more low-energy cosmic rays present in these clouds than is measured by {\it Voyager} in the local interstellar medium. This could be due to low-energy cosmic rays from young stars close to the molecular clouds. However, the stars in our sample are all nearby stars within 7\,pc of the solar system. Thus, adopting the LIS for our sample is a good assumption. 

We also note that, if low energy cosmic rays were more abundant than the LIS it is unlikely to affect the radiation dose calculated in our paper for the surface of GJ~273\,b (see \Cref{sec:radiation}). In addition, close-in exoplanets (such as the ones we have in our sample) would not be affected by low energy cosmic rays because they are suppressed strongly by advective processes. This might not be the case for planets that orbit further out.

\section{Galactic cosmic ray fluxes around M dwarfs}
\label{sec:results}
\subsection{The flux of Galactic cosmic rays at the habitable zone and at planetary orbits}
\label{sec:crs}
The habitable zone depends on the planetary mass and atmospheric conditions. In this work we calculate the habitable zone size using the prescription of \citet{Kopparapu2014}, where the distance in au is given by
\begin{equation}
    d = \sqrt{\frac{L/L_\odot}{S_{\text{eff}}}},
    \label{eq:hz}
\end{equation}
where $L/L_\odot$ is the stellar bolometric luminosity compared with the solar luminosity 
and $S_{\text{eff}}$ is the stellar effective flux incident on the top of the planet's atmosphere, given by
\begin{equation}
    S_{\text{eff}}= S_{\text{eff}\odot} + aT_\star + bT^2_\star + cT^3_\star + dT^4_\star,
    \label{eq:seff}
\end{equation}
where $T_\star= T_{\text{eff}}- 5780\,\text{K}$. The coefficients in \Cref{eq:seff} are given by the recent Venus and early Mars limits in Table 1 of \citet{Kopparapu2014} for planets with $ 0.1\,M_\oplus \le M_{\rm planet} \le 5\,M_\oplus$. The habitable zone boundaries that we calculate are given in  \Cref{tab:stars}. With this prescription, only one exoplanet (GJ~273\,b) lies in the habitable zone, with GJ~887\,c lying very close (at 0.12\,au) to the inner edge of the habitable zone (at 0.16\,au).

\begin{figure*}
    \includegraphics[width=0.86\paperwidth]{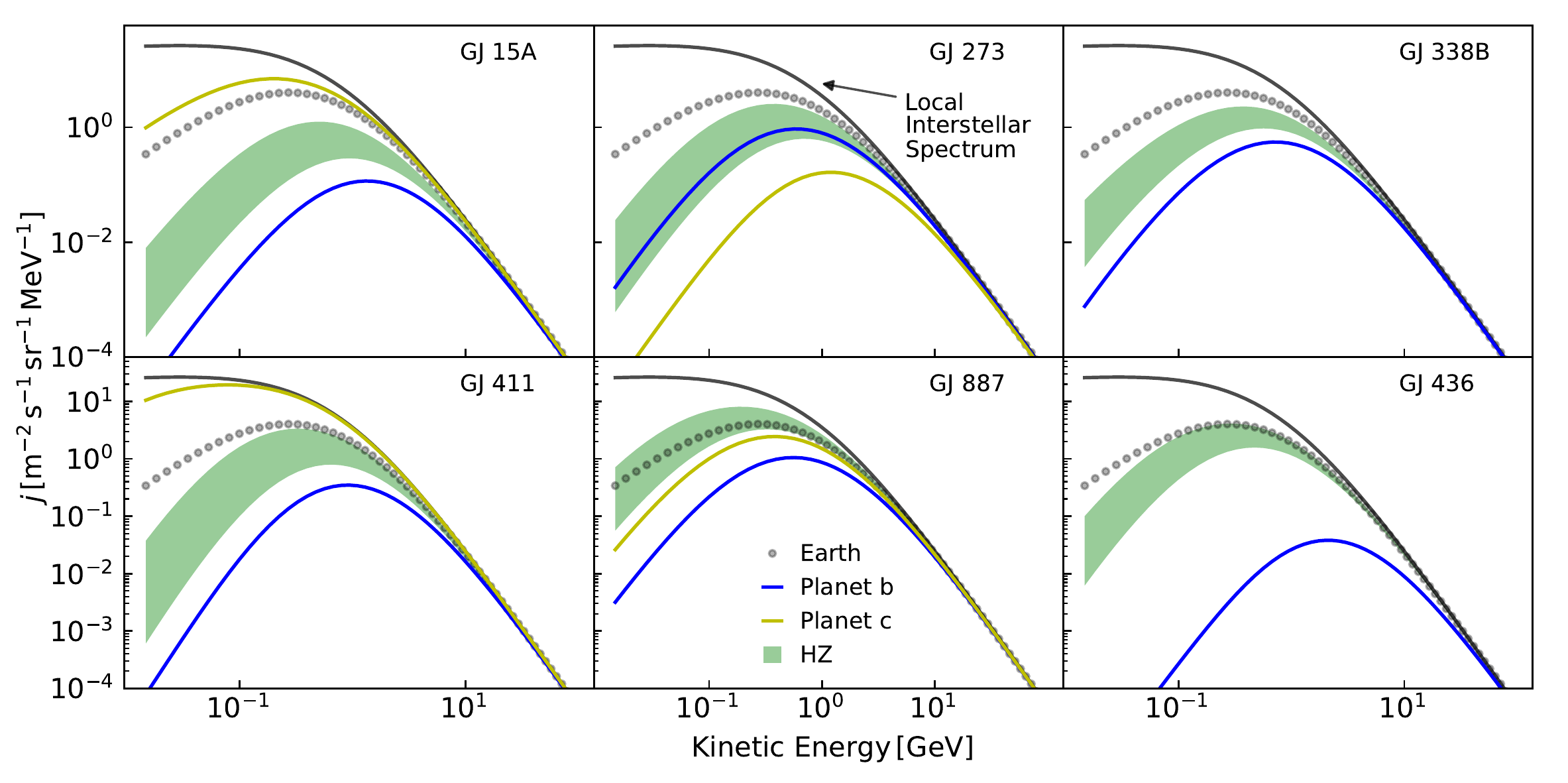}
    \caption{Differential intensity of Galactic cosmic rays as a function of kinetic energy for six M dwarfs. The green shaded areas are the flux of cosmic rays in the habitable zone of each star. The blue and yellow lines are the fluxes of cosmic rays at planets b and c orbital distances, respectively. The grey dots in each panel is the Galactic cosmic rays fluxes observed at Earth and the black line is the LIS.}
    \label{fig:HZs}
\end{figure*}

\Cref{fig:HZs} shows the differential intensity of cosmic rays as a function of kinetic energy in the habitable zone (green shaded areas), at planet b orbit (blue curves) and at planet c orbit (yellow curves) for each star in our sample and GJ~436 \citep[case A from][]{Mesquita2021}. For each panel, the solid black line is the LIS and the grey dots are representative of the Galactic cosmic ray spectrum observed at Earth's orbit, representative of solar minimum values \citep[taken from a model from][]{Rodgers2020}. With the exception of GJ~15A\,c and GJ~411\,c all the planets in our sample receive a lower flux of cosmic rays in comparison with Earth (see \Cref{fig:HZs}) for all kinetic energies because they orbit close-in. In particular, we observe a strong suppression of low-energy cosmic rays at the majority of exoplanets orbit. Galactic cosmic ray fluxes are seen to continue decreasing for close-in distances for the Sun as well \citep[during solar minimum observations from][]{Marquardt2019}. However, the solar observations also capture temporal variations that we do not take into account in our work. For example, our model neglects velocity drift terms and is most applicable to solar minimum conditions (when the velocity drift term is minimal). 

The bottom panels of \Cref{fig:HZs} show that two stars in our sample have Earth-like Galactic cosmic ray fluxes in their habitable zones, namely GJ~887 and GJ~411, although GJ~411 is only comparable for cosmic rays with energies above 0.4\,GeV energies. GJ~436 also has comparable Galactic cosmic ray fluxes at its habitable zone for energies larger than 0.1\,GeV. Interestingly, GJ~273\,b, which is the only exoplanet in our sample in the habitable zone, receives a much lower (up to two orders of magnitude) flux of Galactic cosmic rays when compared with the values at Earth. We will come back to this planet when we calculate the biological surface radiation dose in the next subsection.

\begin{figure}
	\includegraphics[width=\columnwidth]{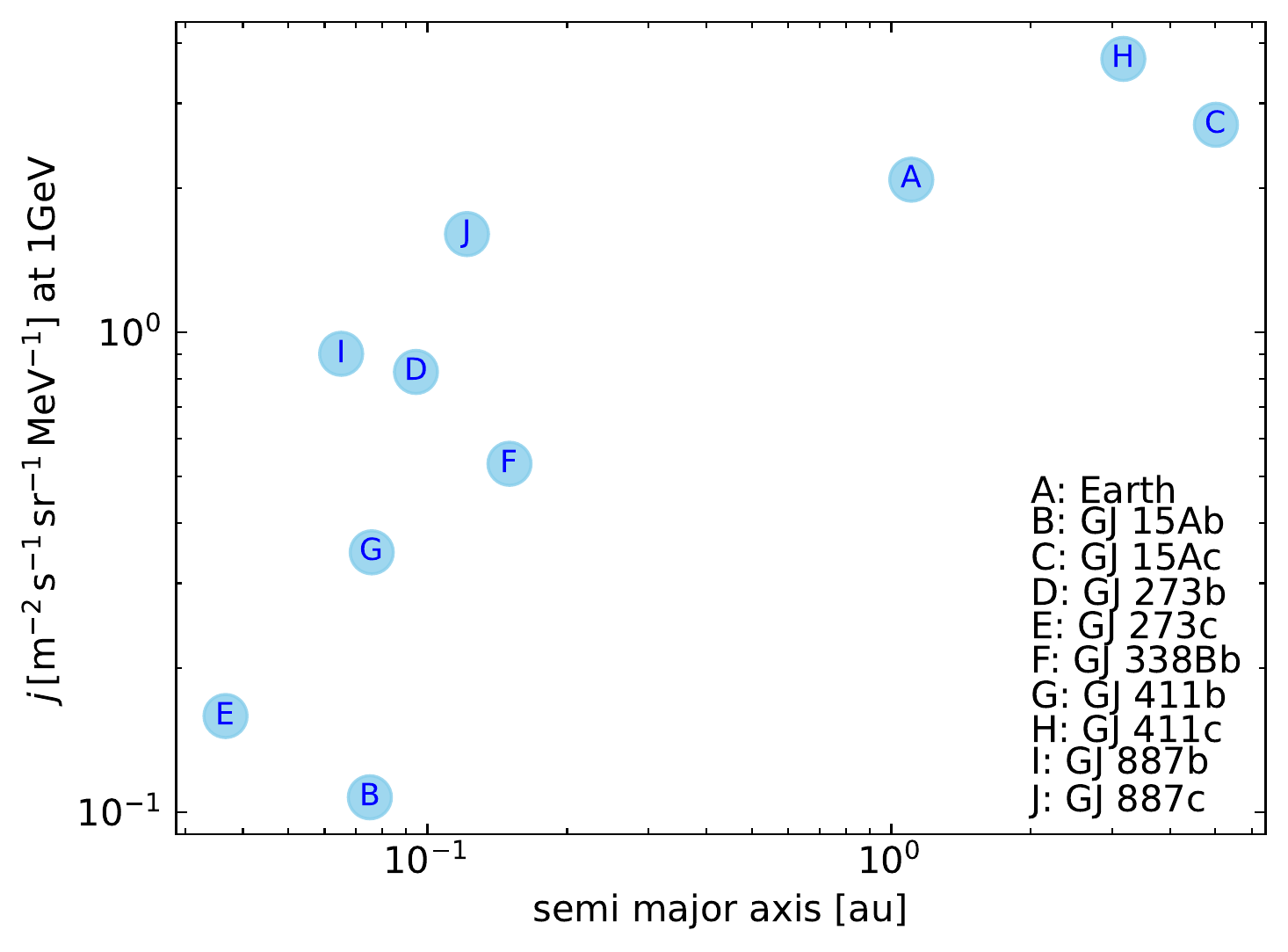}
    \caption{Galactic cosmic ray flux at semi-major axis of each planet for particles with 1\,GeV kinetic energy. Most of the planets orbiting the M dwarfs studied here have a very small semi-major axis with exception of GJ~15Ac and GJ~411\,c.}
    \label{fig:planets}
\end{figure}

\Cref{fig:planets} shows the flux of GeV energy Galactic cosmic rays as a function of semi-major axis for each planet in our sample and the Earth. We chose 1\,GeV energy cosmic rays because particles with this energy can penetrate exoplanetary atmospheres, as they do not lose as much energy as low-energy cosmic rays. Almost all the planets in our sample have a very small semi-major axis, with exception of GJ~15A\,c and GJ~411\,c. For a given star, the Galactic cosmic ray fluxes decrease with decreasing orbital distance. Thus, a closer-in planet receives lower Galactic cosmic rays fluxes in comparison to a planet with a larger semi-major axis. We identify that different wind parameters may coincidentally lead to similar levels of Galactic cosmic rays for different planetary systems (see \Cref{fig:planets} planets GJ~273\,b (D) and GJ~887\,b (H), for instance).

Here, we focused only on Galactic cosmic rays but active stars with strong magnetic fields should be efficient in accelerating energetic particles or stellar cosmic rays \citep{Rodgers2021}. Similarly to Galactic cosmic rays, stellar energetic particles can also interact/affect exoplanets' magnetospheres, atmospheres and surface \citep{Segura2010, Grenfell2012, Tabataba2016, Fraschetti2019, Scheucher2020}. In addition, because M dwarfs have close-in habitable zones and several observed close-in exoplanets, it is expected that at those distances the stellar cosmic rays may dominate up to a given energy.

\subsection{GJ~273\,b's surface radiation dose}
\label{sec:radiation}
GJ~273\,b is a super-Earth orbiting in the habitable zone at a distance of 0.091\,au. Its mass is $2.89\,M_\oplus$ and it has an orbital period of 18.6 days \citep{Astudillo2017}. It receives an incident bolometric flux of $1.06$ times that received at Earth \citep{Astudillo2017}. In addition to the presence of surface liquid water, GJ~273\,b could be potentially habitable if an atmosphere is present. Here, assuming an Earth-like atmosphere, we investigate the biological radiation dose that reaches GJ~273\,b's surface.  

The biological radiation dose is modelled using the GEANT4 package \citep{Agostinelli2003}. It is a Monte Carlo code developed at CERN to model charged particle interactions with matter and is extensively calibrated and used worldwide. For simplicity, we assume that GJ~273\,b has an Earth-like atmosphere and has no global magnetic field. We use the standard Earth's atmosphere as used in the earlier studies \citep{Atri2017, Atri2020} incident with isotropic flux of particles ranging from energies corresponding to our assumed $p_{\rm min}$ and $p_{\rm max}$ values. The radiation dose is calculated on an ICRU (International Commission on Radiation Units and Measurements) sphere-equivalent of 15\,cm radius on the surface of the planet consisting of 100\% water, assuming that life if it exists on the planet it is likely to be water-based. 
We obtain a dose equivalent rate of $4.12\times 10^{-12}$\,Sv/s, which is 0.13\,mSv/yr. For comparison, the annual dose equivalent on the Earth's surface is around 0.33\,mSv (according to the National Council on Radiation Protection and Measurements report No 160). Even though the 15\,MeV Galactic cosmic ray flux at GJ~273\,b is about 200 times smaller than at Earth, the difference in flux is smaller at energies of a few 100\,MeV and above, which is the part that contributes most to the radiation dose at the surface. Lower energy particles deposit energy in the top of the atmosphere, and do not contribute to the radiation dose on the surface.

As mentioned in the \Cref{sec:crs}, active stars can also generate stellar cosmic rays. Stellar cosmic rays with energies $\gtrsim 100$\,MeV may also contribute to the radiation dose at the planetary surface and particles with energies $\lesssim 100$\,MeV can ionise the planet's atmosphere.

\subsection{How are Galactic cosmic rays modulated by the magnetised wind of M dwarfs?}
\begin{figure}
	\includegraphics[width=\columnwidth]{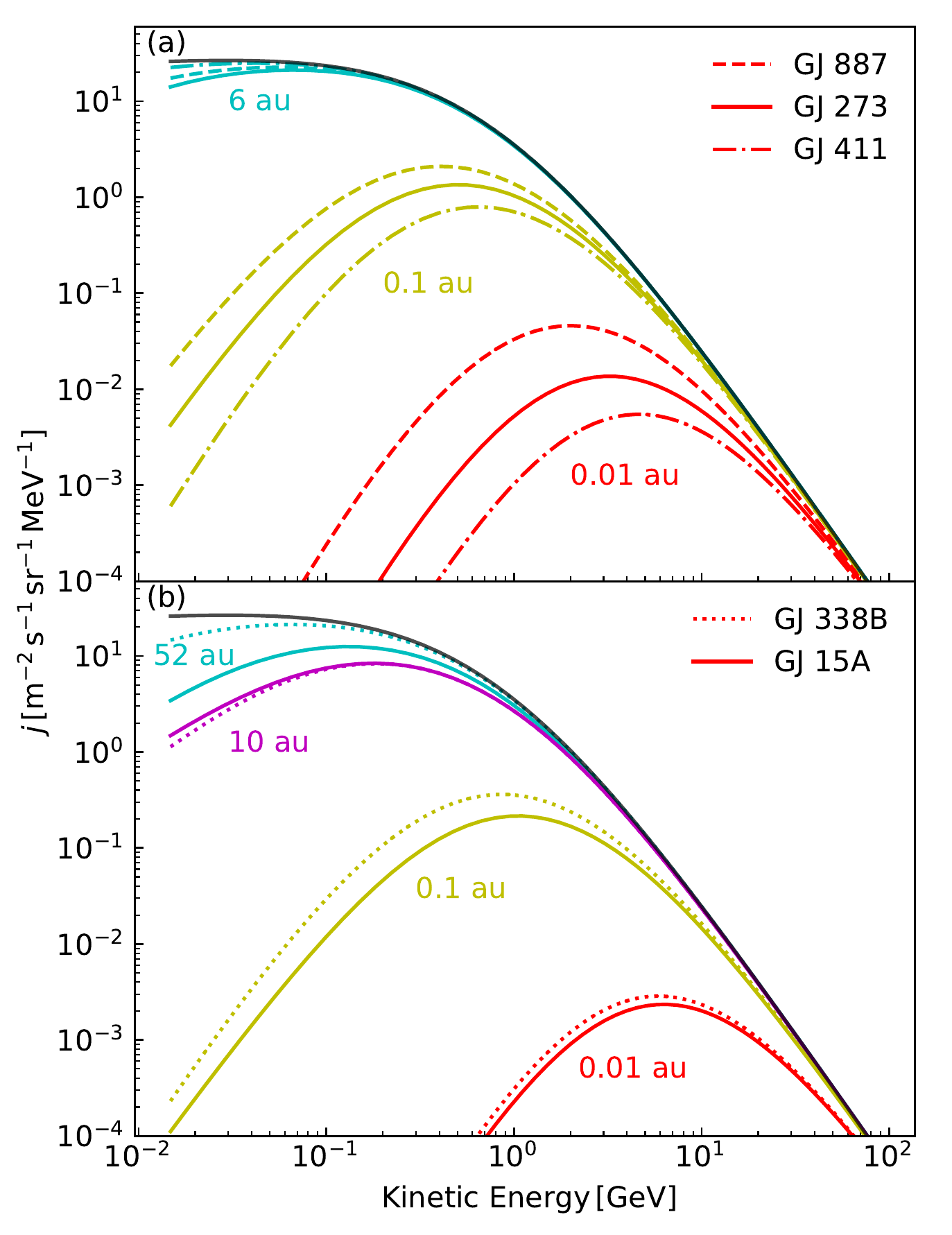}
    \caption{Differential intensity of Galactic cosmic rays as a function of kinetic energy. a) Group 1: solid lines GJ~273, dash-dotted GJ~411 and dashed GJ~887. b) Group 2: solid lines GJ~15A and dotted lines GJ~338B. The colours represent the same distances in each plot and the solid black line is the LIS.}
    \label{fig:cases}
\end{figure}

Here we investigate how Galactic cosmic rays are modulated by the stellar winds of different M dwarfs. We divided our sample of stars in two main group according to their radial magnetic field/rotation period. Group 1 includes: GJ~273, GJ~411 and GJ~887 which have longer rotation periods and smaller radial magnetic field strengths. Group 2 includes: GJ~15A and GJ~338B which have shorter rotation periods and larger radial magnetic field strengths when compared with group 1. \Cref{fig:cases} shows the differential intensity of Galactic cosmic rays as a function of cosmic ray kinetic energy for group 1 (\Cref{fig:cases}a) and group 2 (\Cref{fig:cases}b).

The magnetic field and velocity properties of the stellar wind are important for the propagation of Galactic cosmic rays. In general, a stellar wind with a strong magnetic field can strongly suppress the flux of cosmic rays. This is because a strong magnetic field results in a small diffusion coefficient which leads to the Galactic cosmic rays being strongly suppressed. To a lesser extent, a stellar wind with higher velocity is also efficient in suppressing the propagation of Galactic cosmic rays. 

\Cref{fig:cases}a shows that GJ~887 (dashed lines) has larger cosmic ray fluxes for the same distance in comparison to the other stars in group 1. This occurs because GJ~887 has the weakest magnetic field and, as a consequence, the cosmic rays are not suppressed significantly by its stellar wind. From group 1, GJ~273 (solid lines) has the strongest magnetic field, however, it modulates the Galactic cosmic rays less than GJ~411 (dash-dotted lines) with a smaller magnetic field. This is explained by the higher velocity wind of GJ~411, around 1.6 times higher than GJ~273. This behaviour, however, is not observed at all radii. At around 1\,au for GJ~273 and GJ~411 the fluxes becomes comparable. The distance where it happens is due to a combination of magnetic field strength, the wind velocity and the size of the astrosphere. 

In relation to group 2, GJ~15A (solid lines in \Cref{fig:cases}b) is the one with the strongest radial magnetic field and the largest astrosphere. Naively, one might expect, since GJ~15A has a strong radial magnetic field, a stronger suppression of Galactic cosmic rays in comparison with all other stars for any given distance. However, when compared with GJ~338B (dotted lines), GJ~15A (solid lines) does not always modulate the cosmic rays more as can be observed in \Cref{fig:cases}b at 10\,au, for instance. The explanation for this behaviour lies in the total magnetic field profile of GJ~338B. Because GJ~338B rotates faster (compared with other stars in the sample), its azimuthal magnetic field profile (blue curve of \Cref{fig:wind_results} right panel) has a larger contribution than the radial magnetic field (green curve of \Cref{fig:wind_results} right panel) for distances greater than 2\,au. As a consequence, the flux of cosmic rays becomes comparable for both stars at about 10\,au (pink lines of \Cref{fig:cases}b).

\begin{figure}
	\includegraphics[width=\columnwidth]{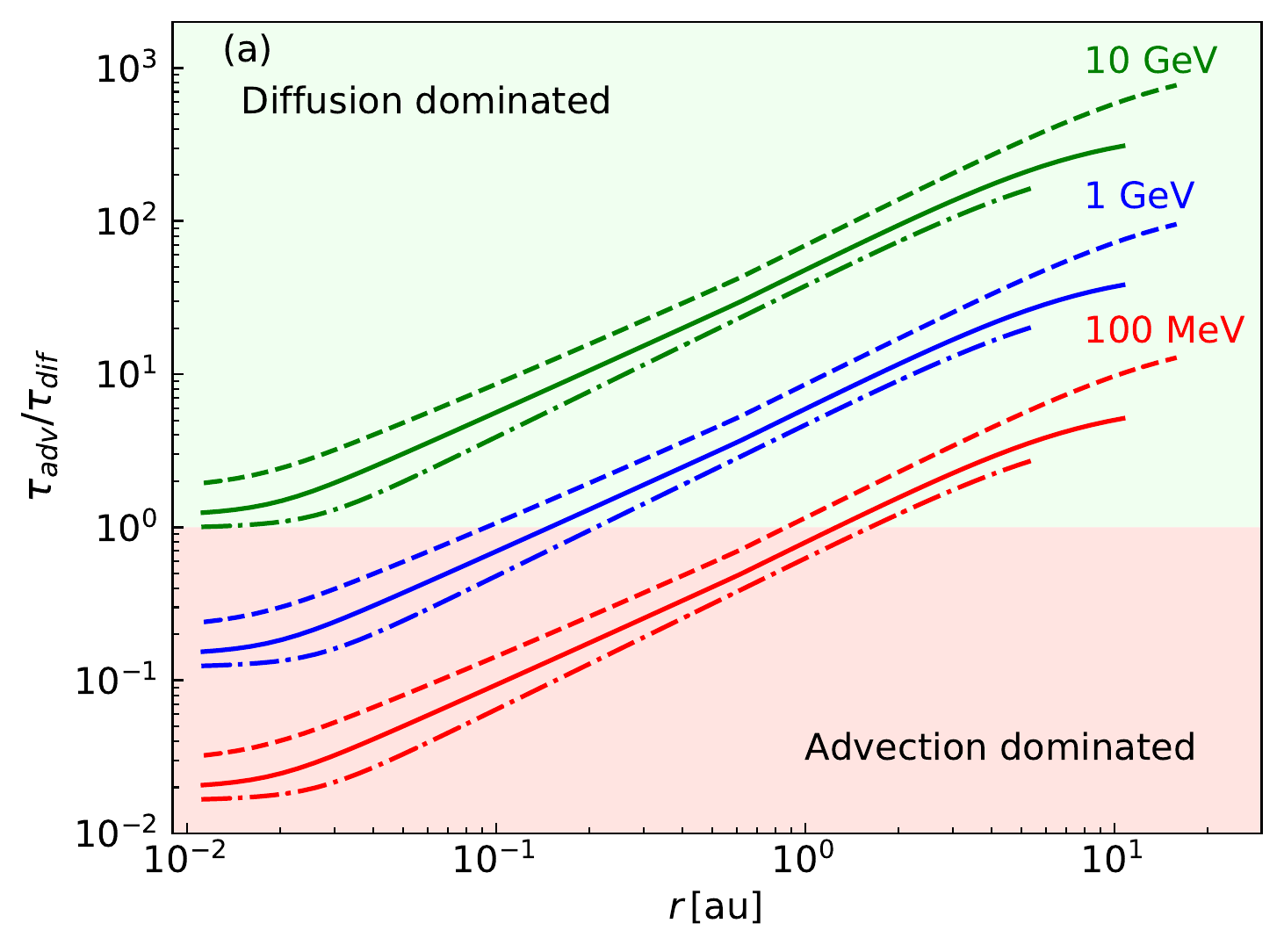}
	\includegraphics[width=\columnwidth]{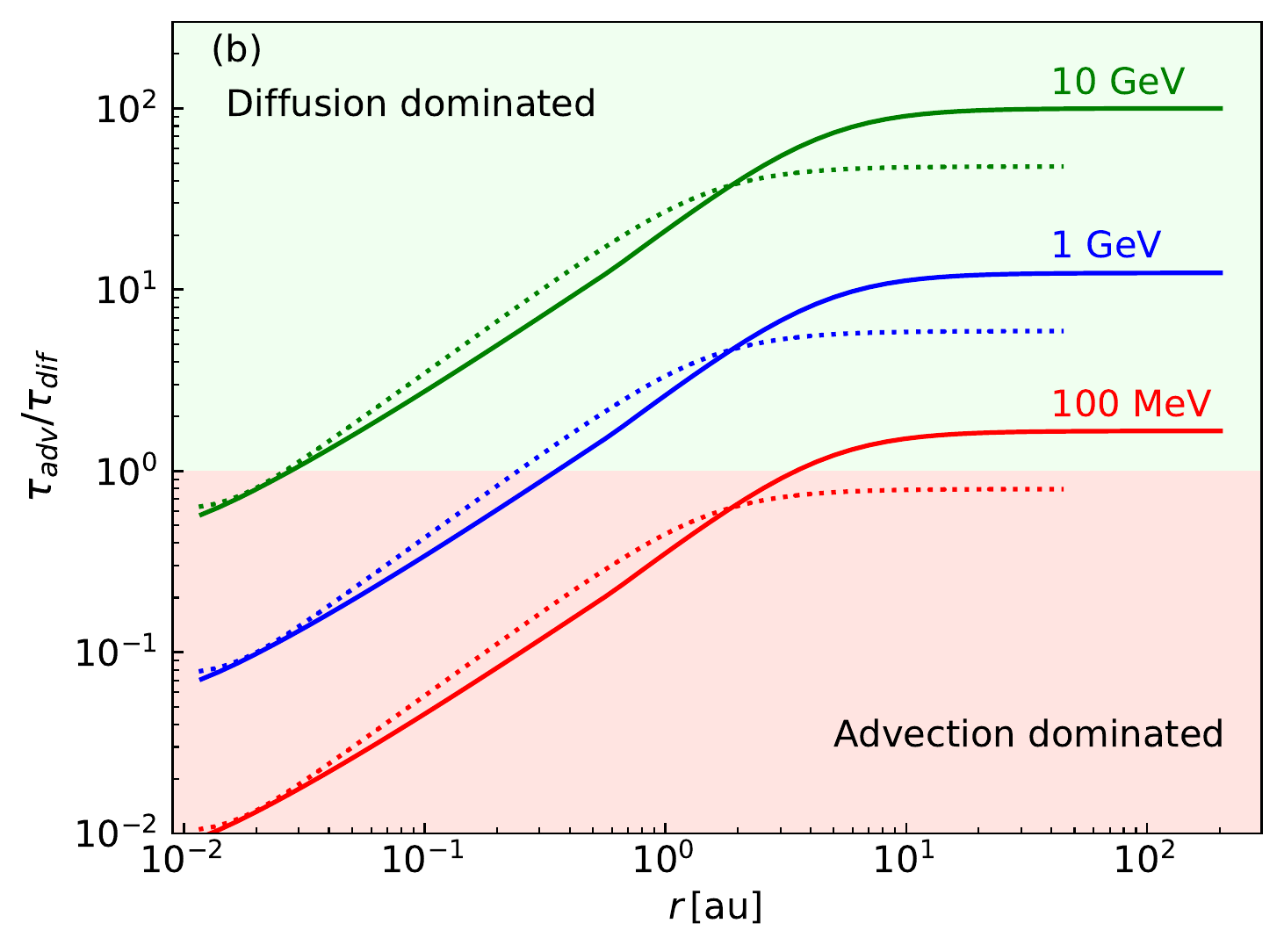}
    \caption{Ratio between advective and diffusive time-scale as a function of distance. The green shaded area is where diffusion dominates and the red shaded area is advection dominated. a) Group 1: solid lines GJ~273, dash-dotted GJ~411 and dashed GJ~887. b) Group 2: solid lines GJ~15A and dotted lines GJ~338B.}
    \label{fig:time-cases}
\end{figure}

A way to understand the modulation of cosmic rays is by investigating the time-scales of the physical process involved in the cosmic ray propagation. The advective and diffusive time-scales are defined as:
\begin{equation}
    \tau_{\text{adv}}=\frac{r}{u}, \qquad\qquad \tau_{\text{dif}}=\frac{r^2}{\kappa(r,p)}\propto \frac{r^2}{p/B}\,.
\end{equation}
\Cref{fig:time-cases} shows the ratio between the advective and diffusive time-scale as a function of the distance for group 1 (\Cref{fig:time-cases}a) and group 2 (\Cref{fig:time-cases}b) for different values of cosmic ray kinetic energy. The red shaded area is where advection dominates ($\tau_{\text{adv}}/\tau_{\text{dif}} <1$) and the cosmic rays are strongly modulated by the stellar wind. The green shaded area is where diffusion dominates ($\tau_{\text{adv}}/\tau_{\text{dif}} >1$) and the cosmic rays experience little to no modulation.

From group 1 (\Cref{fig:time-cases}a), GJ~411 (dash-dotted lines) has the smallest time-scale ratio for all energies and distances and is followed by GJ~273  (solid lines) and GJ~887 (dashed lines). For this reason, GJ~411 modulates the cosmic rays more than the other two stars in group 1. From group 2 (\Cref{fig:time-cases}b), GJ~15A (solid lines) has the smallest time-scale for $r<2$ \,au, and as a consequence, it has lower cosmic ray fluxes. For $r>2$\,au, GJ~338B (dotted lines) has the smallest time-scale but at these distances, diffusion dominates and the cosmic rays suffer little (or no) modulation.

For 1\,GeV energy cosmic rays (\Cref{fig:time-cases} blue curves), the region where the cosmic rays transition from the advection-dominated regime to the diffusive-dominated regime happens at $\lesssim 0.3$\,au for all stars in our sample. This means that for $r\gtrsim 0.3$\,au diffusion dominates and the cosmic rays are not strongly modulated. \citet{Rodgers2021-2} proposed that if diffusion dominates at larger orbital distances varying the size of the astrosphere have almost no effect on the flux of Galactic cosmic rays. This implies that the Galactic cosmic ray spectrum for the stars in our sample should not be strongly affected by variations in astrosphere size as was observed for the case of GJ~338B and GJ~436 \citep{Mesquita2021}.

\section{Discussion \& Conclusions}
\label{sec:conc}
In this paper, we investigated the differential intensity of Galactic cosmic rays within the astrospheres of M dwarfs. We focus on the habitable zones and at the planets orbit. We also investigate the wind properties of the M dwarfs in our sample. Our sample of stars were specially selected to include only M dwarfs with known planets and mass-loss rate measurements from Lyman-$\alpha$ observations. We selected five M dwarfs, namely: GJ~15A, GJ~273, GJ~338B, GJ~411 and GJ~887. These stars each have one or two known exoplanets and in total our sample contains 9 exoplanets including super-Earths and super-Neptunes.

In our simulations, the orbits of GJ~273\,c and GJ~411\,b lie within the Alfv\'{e}n radius, in a sub-Alfv\'{e}nic region. This configuration can potentially cause a star-planet interaction signature on the star as energy can be transported back to the star \citep{Ip2004,Saur2013}. Signatures of such interactions include anomalous CaII H\&K emission \citep{Shkolnik2008,Cauley2019} and planet-induced radio emission \citep{Vedantham2020, Kavanagh2021}.

The stellar wind velocity and magnetic field profiles play an important role in the propagation of Galactic cosmic rays. A stronger stellar magnetic field profile results in a lower flux of Galactic cosmic rays in the astrosphere when compared with a smaller magnetic field strength. To a lesser extend, a stronger stellar wind velocity also results in a lower Galactic cosmic ray fluxes in the astrosphere in comparison with a weak stellar wind velocity. 

The mass-loss rate estimate from Lyman-$\alpha$ observations of the two binary systems in our sample is a combination of the two stars. When simulating the stellar wind we assume that the total mass-loss rate is the individual contribution of a single star in the binary system. This assumption may not be the best approach but how much it would affect the results of the Galactic cosmic ray propagation? If for instance, the mass-loss rate is a contribution of 80\% star A and 20\% star B it probably means that the stellar properties of star A and B are different. If the magnetic field and the stellar wind velocity are different for star A and B it would probably cause an effect on the Galactic cosmic ray fluxes in each astrosphere. 

Two stars in our sample have an Earth-like level of Galactic cosmic rays in their habitable zone, namely GJ~411 and GJ~887 \citep[similar to what was found for GJ~436 by][]{Mesquita2021}. GJ~15A, GJ~273 and GJ~338B have lower Galactic cosmic rays fluxes in their habitable zone in comparison with Earth. 

GJ~273\,b is the only known exoplanet in the habitable zone in our sample. However, it receives a much lower Galactic cosmic ray flux than Earth, up to two orders of magnitude for 15\,MeV energy cosmic rays. In addition, GJ~887\,c lies close to the inner edge of the habitable zone, and its Galactic cosmic ray flux is around 10 times lower than Earth's value at 15\,MeV cosmic ray energies. The other planets in our sample, with the exception of GJ~15A\,c and GJ~411\,c, show a higher suppression of Galactic cosmic rays when compared with Earth because they orbit much closer-in. Opposite to the other planets in our sample, GJ~15A\,c, has a larger semi-major axis and it receives slightly higher Galactic cosmic ray fluxes than Earth. GJ~411\,c, which has also a larger semi-major axis, receives a much higher flux of cosmic rays (comparable with the LIS values) as it orbits close to the outer edge of GJ~411's astrosphere. Interestingly, due to its close proximity to the astrosphere edge, GJ~411\,c atmosphere could be affected by an enhancement of low energy cosmic rays in the LIS. Depending on the temperature of the planet, GJ~411\,c may be a good candidate to study the impact of Galactic cosmic rays on atmospheric chemistry. Spectroscopic observations of molecular features from ions, such as H$_3$O$^+$ and NH$_4^+$ \citep{Helling2016,Barth2020}, with the James Webb Space Telescope \citep[JWST,][]{Gardner2006} and the Atmospheric Remote-sensing Infrared Exoplanet Large-survey \citep[Ariel,][]{Tinetti2021} could possibly constrain the incident cosmic ray spectrum and detect the existence of a possible excess of low-energy particles. 

In our sample, the propagation of Galactic cosmic rays at large radii is dominated by diffusion, and according to \citet{Rodgers2021-2} a change in the astrosphere size for this type of system does not strongly affects the spectrum of Galactic cosmic rays. This is what we observe for GJ~338B when we increased the astrosphere by 60\%. For systems dominated by diffusion, thus, our lack of knowledge for the ISM properties does not strongly affect the Galactic cosmic ray propagation. To determine if the propagation of Galactic cosmic rays in a system is dominated by diffusion or advection it is necessary to know the stellar wind velocity and magnetic field. However, it does not require knowledge of the ISM properties.

It is possible to quantify the impact of cosmic rays on life-forms by calculating the radiation dose a planet receives on its surface. Assuming GJ~273\,b has an Earth-like atmosphere and no magnetic field we estimated that it receives an equivalent dose of 0.13\,mSv/yr. This value is around 40\% of the annual dose on Earth's surface. Although GJ~273\,b receives two orders of magnitude less 15\,MeV energy cosmic rays than Earth, for high energy particles ($\sim$\,GeV) the difference in fluxes are much smaller (2.3 times less). That is the reason why the radiation dose on GJ~273\,b's surface is quite significant, because high-energy particles ($\gtrsim 100$\,MeV) contribute most to the radiation dose at the planet's surface.

What are the implications of a star having a similar Galactic cosmic ray flux as observed at Earth in their habitable zone? The level of Galactic cosmic rays Earth receives is not harmful for life as we know it. In comparison, the propagation of Galactic cosmic rays to Earth when life is thought to have started results in a significant reduction of Galactic cosmic rays in comparison with the present-day Earth values \citep{Cohen2012, Rodgers2020}. If life already exists on those planets it would not be negatively affected by the effects of Galactic cosmic rays. These assumptions also depend on whether the planet has an atmosphere and/or a magnetic field. If life still does not exist on those planets the Galactic cosmic ray fluxes may be important for the start of life \citep{Rimmer2014, Airapetian2016}.

The Parker Solar Probe will be able to measure the Galactic cosmic ray spectrum in the inner heliosphere \citep{Marquardt2019}, which will help to better characterise cosmic ray models for close-in exoplanets. A 2D (or 3D) cosmic ray transport model could be used in the future to more accurately model Galactic cosmic ray propagation, as is commonly used for the solar system \citep{Potgiete2015}. However, given the lack of observational constraints for the type and level of turbulence in M dwarf winds 1D models seem well-motivated currently. The results found here can be further used to investigate the Galactic cosmic ray fluxes at the magnetospheres and atmospheres of the exoplanets in our sample. 

\section*{Acknowledgements}
This project has received funding from the European Research Council (ERC) under the European Union's Horizon 2020 research and innovation programme (grant agreement No 817540, ASTROFLOW). The authors wish to acknowledge the SFI/HEA Irish Centre for High-End Computing (ICHEC) for the provision of computational facilities and support. ALM and AAV acknowledge funding from the Provost's PhD Project award. DA acknowledges support from the New York University Abu Dhabi (NYUAD) Institute research grant G1502.

\section*{Data Availability}
The data described in this article will be shared on reasonable request to the corresponding author.



\bibliographystyle{mnras}
\bibliography{bib/reference.bib}





\bsp	
\label{lastpage}
\end{document}